# Micro-pulse upconversion Doppler lidar for wind and visibility detection in the atmospheric boundary layer

Haiyun Xia,[1,3] Mingjia Shangguan,[1,2] Chong Wang,[1] Guoliang Shentu,[2,4] Jiawei Qiu,[1] Qiang Zhang,[2,4,5] Xiankang Dou,[1,*] Jianwei Pan[2,4]

[1] CAS Key Laboratory of Geospace Environment, University of Science and Technology of China, Hefei, 230026, China

[2] Shanghai Branch, National Laboratory for Physical Sciences at Microscale and Department of Modern Physics, USTC, Shanghai, 201315, China

[3] Collaborative Innovation Center of Astronautical Science and Technology, Harbin Institute of Technology, Harbin 150001, China

[4] Synergetic Innovation Center of Quantum Information and Quantum Physics, USTC, Hefei 230026, China

[5] Jinan Institute of Quantum Technology, Jinan, Shandong 250101, China

*Corresponding author: dou@ustc.edu.cn

## Abstract

For the first time, a versatile, eyesafe, compact and direct detection Doppler lidar is developed using upconversion single-photon detection method. An all-fiber and polarization maintaining architecture is realized to guarantee the high optical coupling efficiency and the system stability. Using integrated-optic components, the conservation of etendue of the optical receiver is achieved by manufacturing a fiber-coupled periodically poled Lithium niobate waveguide and an all-fiber Fabry-Perot interferometer (FPI). The so-called double-edge direct detection is implemented using a single-channel FPI and a single upconversion detector, incorporating time-division multiplexing method. The relative error of the system is lower than 0.1% over 9 weeks. To show the robust of the system, atmospheric wind and visibility over 48 hours are detected in the boundary layer. In the intercomparison experiments, lidar shows good agreement with the ultrasonic wind sensor (Vaisala windcap WMT52), with standard deviation of 1.04 m/s in speed and 12.3° in direction.

## 1. Introduction

As the atmospheric boundary layer (ABL) affects both the dynamics and thermodynamics of the atmosphere, it plays an important role in many fields, such as air pollution and the dispersal of pollutants, agricultural meteorology, hydrology, numerical weather prediction, climate simulation and aeronautical meteorology. In air quality control and management, to understand the transport and dispersal processes that affect concentrations of atmospheric pollutants, one needs knowledge of dynamic and mixing conditions including wind speed and direction profiles, strength of turbulence, and structure of the ABL [1]. In aeronautical applications, detections of boundary-layer phenomena such as low cloud and fog affecting atmospheric visibility, microbursts and low-level jets leading to wind shear and clear air turbulence, are of crucial importance for flight safety and airport capacity [2, 3].

Various remote sensing instruments such as sodars (acoustic sounders); acoustic radars; radars and lidars have been developed for atmospheric wind observations. The ability and advantage of Doppler lidars that make continuous, real-time 3D wind detection via Mie or Rayleigh backscatter with high temporal and spatial resolutions distinguishes this technology from other atmospheric sensing technologies [4]. In particular, coherent lidars have been demonstrated successfully on ground-based [5, 6], ship-based [7] and airborne [8-10] platforms. In contrast to earlier coherent lidar based on free-space optics [11], the use of fiber-optic elements and integrated devices at working wavelengths of 1.5 or 2.0 $\mu m$ offers practical advantages including mechanical decoupling and remote installation of the subsystems, simplification of the configuration and alignment, and improvement of the coupling efficiency and long-term stability.

In spite the great success in different applications, there are still challenges in developing the next-generation coherent lidars. First of all, it would be a great challenge to achieve a real-time data processing in the coherent lidar, especially when shorter pulses (requiring higher AOM frequency shift and thus higher sampling rate) are used to obtain a better range resolution. The computation of fast Fourier transform (FFT) and peak detection in the spectral analyses are the stepping stones for the wind estimation. Typically, the averaging number of monitoring pulses is about 50,000, and the number of range bins is larger than 100. For retrieval of one radial wind profile, the peak frequency should be estimated after FFT manipulation of the raw analog signal (for example, 750 MHz sampling rate and 8-bit A/D resolution is used to achieve a range resolution of 15 m, [12]) for 5 mega times, and then averaged in all range bins. On the contrary, in the direct-detection Doppler lidar, backscattering signals can be simply recorded and averaged on a multiscaler (multiple-event time digitizer) in photon-counting acquisition model (2-bit) [13, 14]. Secondly, due to strong dependence of the Rayleigh backscatter cross-section on wavelength ($\beta_R \propto \lambda^{-4}$) and the low sensitivity of coherent detection to spectrally broadened signals (~GHz), the wildly used Klett-Fernald inversion algorithm cannot be used to calibrate the coherent lidar for quantitative aerosol analyses [15]. Finally, the most valuable feature of direct-detection Doppler lidars, relates to the potential for noiseless detection, through using detectors with low-noise and high quantum efficiency [4]. As demonstrated in this letter, the background excess noise is substantially suppressed, whereas, for coherent receivers, shot-noise of local oscillator always exists, even in the absence of atmospheric backscatter.

## 2. Experiment

Now, it is well established that a fiber architecture is easy to adjust and mechanically reliable in a harsh environment. The other advantages of all-fiber architecture are their compactness and flexibility in terms of installation. Using all-fiber architecture as in coherent lidars at communication band, a micro-pulse and versatile direct-detection Doppler lidar is developed for wind and aerosol detection in the ABL. Double-edge technique is realized by using a single-channel all-fiber Fabry-Perot interferometer (FPI) and a single upconversion detector. The lidar can be divided into five subsystems (laser, circulator, telescope, scanner and receiver), linked together using polarization maintaining fiber(PMF), as illustrated in Fig. 1.

The laser adopts master oscillator power amplifier structure. The continuous-wave laser from the distributed feedback diode (DFB) is chopped into pulse train by using two Lithium niobate intensity modulators (Photline, MXER-LN-10) in cascade, which suppresses the CW leakage to -70dB and minimizes the amplified spontaneous emission (ASE) in the erbium-doped fiber amplifier (EDFA). Since only electro-optic modulators (EOM) are used, there is no RF driver for acousto-optic modulator (AOM must be used for frequency shift in coherent lidars). The ASE is further reduced by inserting a fiber Bragg grating (FBG) (AOS GmbH, C-band) with ultranarrow band of 6 pm. Thus the output spectrum of the updated laser is purified as shown in fig. 2, where the pulse energy/duration is set to 50 $\mu J$ /200 ns. In contrast with a commercial available laser (Keyopsys, FEFA-EOKA), the updated laser has a 20 dB lower ASE noise. The circulator is built up with a pair of Brewster plates (BP) and a piece of quarter-wave plate (QWP). Orthogonal staring observation with constant zenith is performed via a telescope and a dual reflection scanner.

The atmospheric backscatter, combined with the reference pulse split out from the laser is fed to the optical receiver. An all-fiber, lensless, FPI (Micron Optics, C-band) is used as the frequency discriminator. The cavity is formed by two highly reflective multilayer mirrors that are deposited directly onto two carefully aligned optical fiber ends. An anti-reflection coated fiber inserted in the cavity provides appropriate confined light-guiding. And a stacked piezoelectric transducer (PZT) is used to axially strain the single-mode fiber inserted in the cavity. Thus, frequency scanning of the FPI can be achieved by changing the cavity length, as we introduced in the earlier direct detection Doppler lidars [13, 14].

The transmitted signal through the FPI is coupled into an upconversion detector (UPD), while the reflected signal is timely delayed, after propagating through a circulator ($C_2$) and an 8-km polarization maintaining fiber (PMF). By using an optical switch (OS, Agiltron, NS-2x2), the transmitted and reflected signals can be directed into the detector alternatively, incorporating time-division multiplexing (TDM) technique.

As shown in Fig. 1, inserts show pictures of the FPI and the UPD. The lidar adopts a polarization maintaining structure, except that the FPI is made of single-mode fiber. However, this can be compensated by adding two polarization controllers (PC) at the front and rear ends of the FPI. The principle and manufacture of the UPD has been introduced elsewhere recently [16-18]. Here, the UPD is integrated into an all-fiber module, in which the periodically poled lithium niobate wave-guide (PPLN-W) is coupled into a PMF/multi-mode fiber (MMF) at the front/rear end. An in-line interferometric filter with 1 nm bandwidth is inserted between the wave-guide and the Si: APD. The quantum efficiency is 20% with a dark noise of 300 counts per second at 1.5 $\mu m$. We emphasize that, due to the lowest attenuation (<0.3dB/km) at 1.5 $\mu m$ in fiber, an all-fiber FPI can be manufactured with very low insert loss (<0.3 dB). The all-solid structure of the FPI makes it immune to ambient fluctuation of atmospheric pressure. And the TDM technique can only be implemented at communication band. In other words, this is the most compact direct detection lidar, as far as we know.

In this work, the response function is defined as,

$$Q(\nu) = [T(\nu) - R(\nu)]/[T(\nu) + R(\nu)], \quad (1)$$

where, $T(\nu)$ and $R(\nu)$ are the transmission and reflection of backscatter on the FPI, respectively.

As shown in Fig. 1, a small fraction of energy of the output laser is split out as the reference signal. By scanning the voltage fed to the PZT in the FPI, the transmission and reflection curves are measured, as shown in Fig. 3(a). And the response function is calculated. To test the stability of the integrated lidar, the response function is measured over 9 weeks, as shown in Fig. 3(b). By applying Voigt fitting to the raw signal, the average full width at half maximum (FWHM) is estimated to be 97.6 MHz with a relative error less than 0.1%. With its novel architecture, in addition to the full use of the backscatter, the simplicity and stability is obtained relative to the conventional direct detection lidar. One should note that, the response value changes from -1 to 1, indicating a doubled sensitivity relative to our Doppler lidar built ten years ago [13].

The lidar can be operated in wind detection mode and aerosol detection mode, depending on the locking point of the laser relative to the FPI, as shown in Fig. 3(a). In the wind mode, the frequency of the output laser is locked at the cross-point, where the sensitivity is maximized. A minor Doppler shift of the backscatter will cause large difference in the response function. In the aerosol mode, the laser is locked at the full-reflection point. And the reflected backscatter is used for quantitative aerosol analyses.

The pulse repetition frequency (PRF=12 kHz) of the laser implies a maximum unambiguous detection range of 12.5 km. The laser beam is pointed at four orthogonal azimuths in sequence with a constant zenith angle of 30°. The dwell time at each azimuth is 10s. Taking the time for scanner movement and data processing into account, each radial wind detection costs 12.5 seconds. The wind speed and direction are calculated based on the assumption of a horizontally homogeneous wind field, using four orthogonal radial wind profiles. Thus, the wind detection period is 50 s. To demonstrate the ability of aerosol detection, after each circle of wind detection, the scanner is point horizontally to north with 6 seconds dwell time for atmospheric visibility detection.

Different from the continuous horizontal backscatter, when the lidar detect wind at zenith angle of 30°, the raw signals drop suddenly at the altitude about 2.5 km, as shown in Fig. 4. This phenomenon is due to the sharp decrease in aerosol concentration at the top of the ABL.

To test the stability of the lidar system, a continuous observation of the atmospheric wind and visibility is started at 12:00 on April 29, and ended at 12:00 on May 1, 2016. The experiment results are shown in Fig. 5. The temperature and humidity near the ground are also monitored for reference. The features of the ABL, such as the wind profiles and depth of the layer, evolve continuously in response to the diurnal cycle of surface heating and cooling. Following sunrise on Apr. 30, a convective layer develops and grows through the morning, reaching a height near 1.6 km by midafternoon. Within the convective ABL, convection transports the heat to the capping inversion base, making the depth of the ABL in accordance with the temperature near the ground. However, after the sunset, wind grows stronger with large gradient, and the depth of a stratified ABL is elevated to a maximum of about 2.5 km. The experiment is stopped just before a thunder shower occurs at 13:30. A low-level jet is observed with maximum speed of 7.8 m/s, at altitude about 0.9 km, at 10:30 on May 1, 2016.

For intercomparison purpose, an ultrasonic wind sensor (Vaisala windcap WMT52) is installed on a tower on the top of our building (31.843°N, 117.265°E), with a height of 54 m above the ground. The temporal resolution is set to 1 minute. The wind speed/direction accuracy is claimed to be ±3% at 10m/s and ±3°, respectively. In total, 2880 detection results are derived, as plotted in Fig. 6. Goods agreements are observed. Histograms of the detection differences between the sensor and lidar are also plotted. The mean differences of wind speed and direction are 0.05m/s and -0.84°, and the standard deviations are 1.04 m/s and 12.3°.

## 3. Conclusion

In conclusion, a most compact direct detection Doppler lidar is demonstrated for continuous wind and aerosol observations in the ABL, incorporating upconversion technique, double-edge technique and time-division multiplexing technique. Since the lidar adopts all-fiber and polarization maintaining architecture, and uses a single-channel Fabry-Perot interferometer and only one upconversion single-photon detector, its system accuracy and stability are improved substantially. As an aeronautical application, for example, wind and visibility are observed continuously over 48 hours. In comparison experiment, good agreements are achieved between the lidar and the ultrasonic wind sensor.

## Acknowledgment

We thank Dr. Mingyang Zheng and Dr. Xiuping Xie from Jinan Institute of Quantum Technology, for their help in manufacturing the upconversion detector.

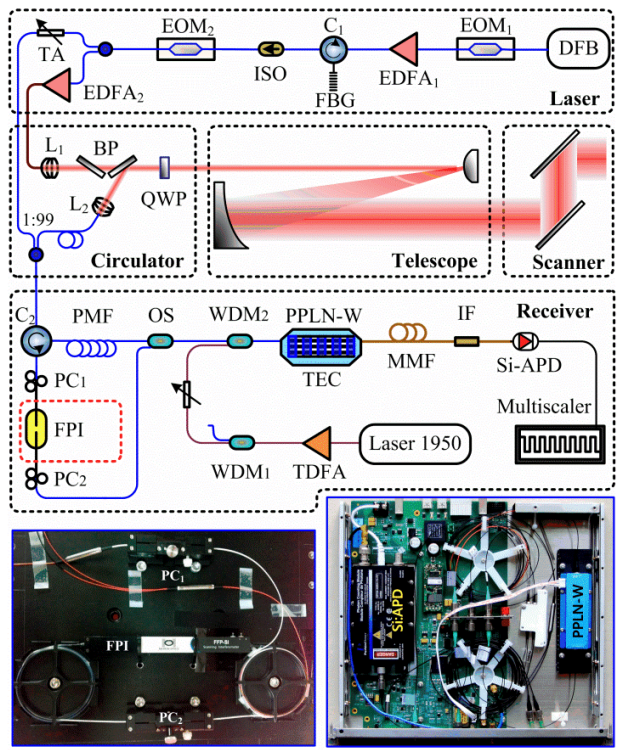

Fig. 1. Schematic of the compact Doppler lidar. TA, tunable attenuator; WDM, wavelength division multiplexer; L, lens; TDFA, thulium-doped fiber amplifier; TEC, thermos-electric cooler.

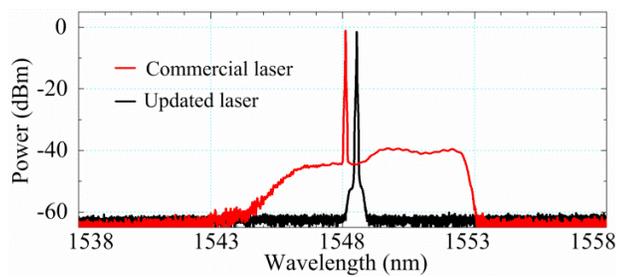

Fig. 2. Output spectra of the lasers. Note the spectrum of the commercial laser is left-shifted by 0.5 nm for comparison purpose.

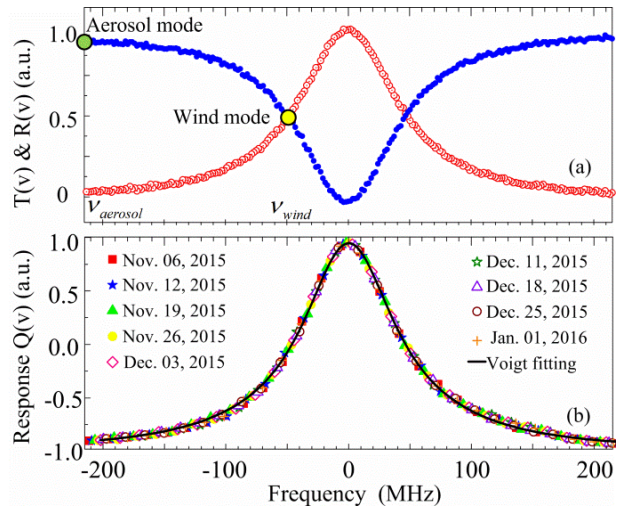

Fig. 3. (a) Transmission and reflection curves, (b) the response functions measured over 9 weeks and one typical Voigt fitting curve.

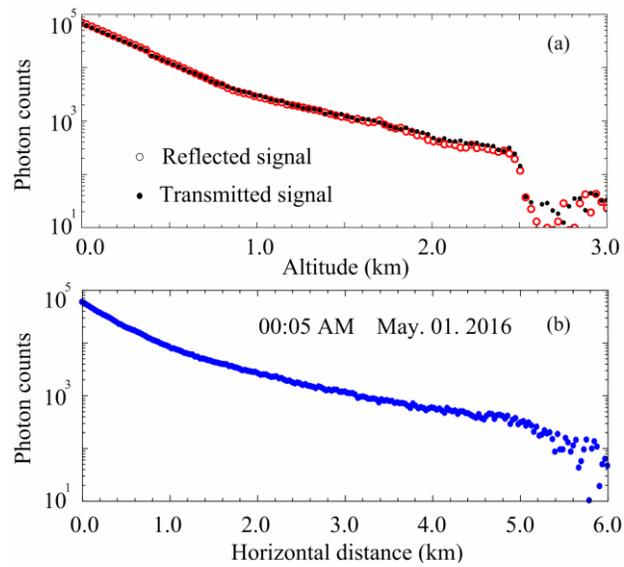

Fig. 4. (a) Reflected and transmitted signals in wind mode, and (b) backscatter signal in aerosol mode.

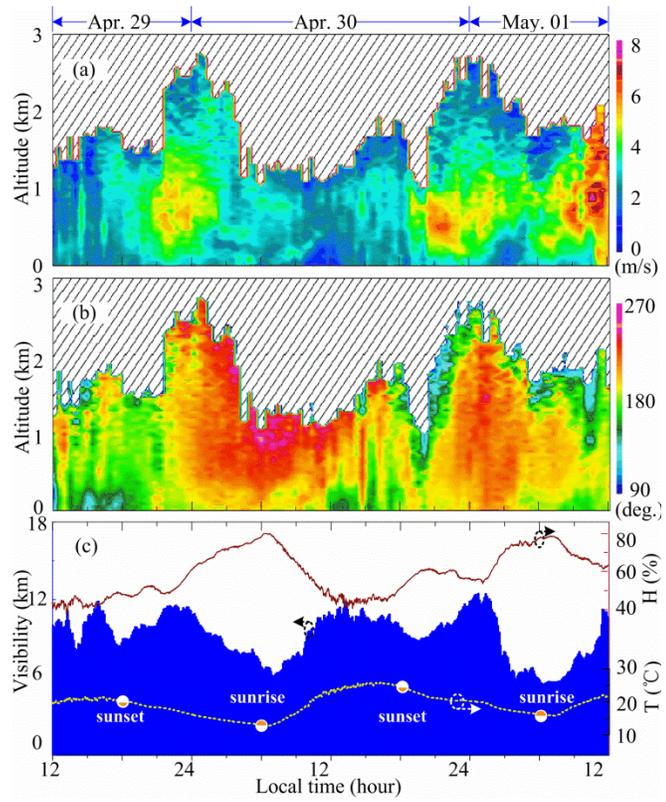

Fig. 5. 48-hour observation of atmospheric wind and visibility. (a) wind speed, (b) direction, (c)visibility, temperature and humidity.

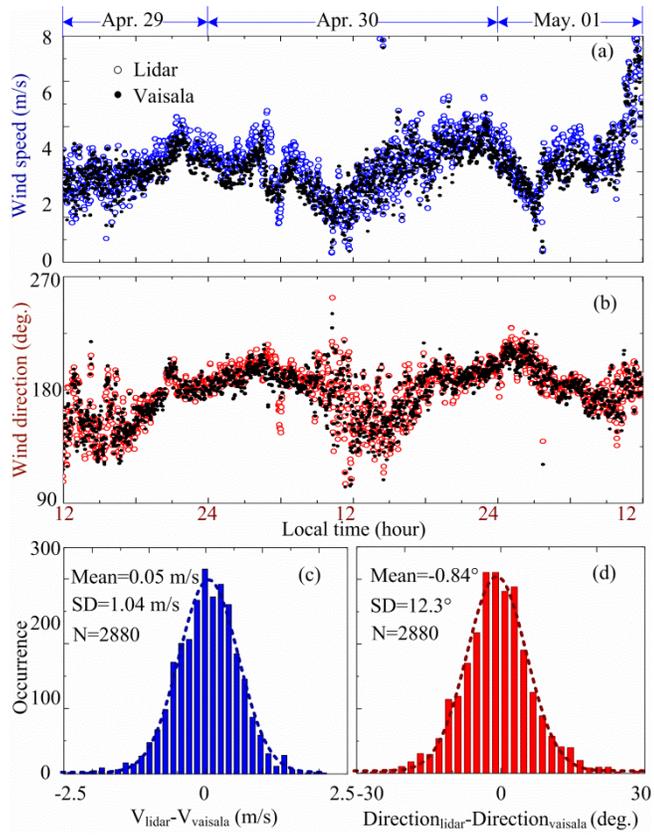

Fig. 6. Comparison experiment results. (a) Wind speed, (b) Direction, and histograms of the differences of speed (c) and direction (d).